\documentclass[prl,twocolumn,showpacs,lengthcheck]{revtex4}

\usepackage{amsmath}
\usepackage{amssymb}
\usepackage{amsthm}
\usepackage{graphicx}

\newcommand{\tr}{\operatorname{tr}}
\newcommand{\uinvnorm}{|\kern-1pt|\kern-1pt|}

\theoremstyle{plain}

\theoremstyle{definition}

\theoremstyle{remark}

\begin{document}
\bibliographystyle{apsrev}

%Title of paper
\title{Lower Bound for the Ground-State Energy Density of a 1D Quantum Spin System}

\author{Tobias J.\ Osborne}
\email[]{Tobias.Osborne@rhul.ac.uk} \affiliation{Department of
Mathematics, Royal Holloway University of London, Egham, Surrey TW20
0EX, UK}

\date{\today}

\begin{abstract}
We present a simple method to calculate systematic lower bounds
for the ground-state energy density of a 1D quantum spin system.
\end{abstract}


\maketitle

The ground-state energy is a quantity of fundamental interest in
the study of quantum spin systems. For example, nonanalytic
behaviour of the ground-state energy as a parameter in the
hamiltonian is varied is a canonical signature of a quantum phase
transition \cite{sachdev:1999a}. The ground-state energy also
occupies a central role in quantum computational complexity
theory: the problem of calculating a good approximation to the
ground-state energy for a class of $2$D spin systems is complete
for the complexity class {\sf QMA}, which is the quantum analogue
of {\sf NP} \cite{kitaev:2002a, kempe:2004a, oliveira:2005a}.

Obviously, because of the {\sf QMA}-completeness results of Kempe
et.\ al.\ \cite{kempe:2004a} and Oliveira and Terhal
\cite{oliveira:2005a} we expect that the ground-state energy is,
in general, extremely difficult to approximate. However, as these
results only pertain to highly disordered/frustrated $2$D systems,
there is some hope that it might be possible to efficiently
calculate approximations to the ground-state energy for $1$D
systems and regular $2$D systems.

This hope has been partially vindicated by the development of the
\emph{density matrix renormalisation group} (DMRG) (see
\cite{schollwoeck:2005a} and references therein for a description
of the DMRG and relatives). The DMRG provides an apparently
efficient algorithm to calculate the ground-state energy and other
local ground-state properties for $1$D quantum spin systems. An
exciting extension of the DMRG to $2$D quantum spin systems was
recently developed by \cite{verstraete:2004a}.

Unfortunately the DMRG cannot \emph{certify} that the estimate it
provides for the ground-state energy is close to the real value. The
reason is that there is no way to rule out the possibility that the
DMRG has become stuck in a local minima. For this reason, if a
promise that the calculated ground-state energy is close to the
correct value is required, then it is vital to either develop
algorithms which provide an estimate on the distance from optima,
or, because the DMRG estimates are always \emph{upper bounds},
provide a method to calculate systematic \emph{lower bounds} for the
ground-state energy.

There has been some previous work on lower bounds for the
ground-state energy: the most important general method which has
been proposed so far is due to Anderson \cite{anderson:1951a}. This
method has been developed further and applied to many situations,
see eg.\ \cite{wittmann:1993a}. The Anderson bound, while
nontrivial, is suboptimal; as we'll see later there exist systems
for which the Anderson bound cannot be systematically improved fast
enough to be useful practically.

In this Letter we describe a simple method to calculate systematic
\emph{lower bounds} for the ground-state energy density (the
ground-state energy per particle) of a $1$D spin system. For general
local $1$D systems it is an efficient numerical procedure to extract
the lower bound. For some systems our method is purely analytic. As
a test we apply our method to calculate a lower bound for
ground-state energy density of the $XY$ model. We also apply our
method to find a lower bound for the average ground-state energy
density of the disordered heisenberg model. Finally, we construct
systems for which the Anderson bound cannot be improved efficiently,
yet our method provides the exact answer.

We consider quantum systems defined on a set of vertices $V$ with a
finite dimensional Hilbert space $\mathcal{H}_x$, i.e.\ a quantum
spin, attached to each vertex $x\in V$. We always assume that $V$ is
finite owing to the standard difficulties \cite{bratteli:1987a,
bratteli:1997a} in defining a thermal state for infinite quantum
spin systems. While we take the limit $n\rightarrow\infty$ we
understand that this limit is purely formal and, strictly speaking,
our results pertain only to the situation where $n$ is large but
finite.

We will, for the sake of clarity, introduce and describe our
results for a ring $\mathcal{C}$ of $n$ distinguishable
spin-$\frac{1}{2}$ particles. Thus, the Hilbert space
$\mathcal{H}_{\mathcal{C}}$ for our system is given by
$\mathcal{H}_{\mathcal{C}} = \bigotimes_{j=0}^{n-1} \mathbb{C}^2$.

We now introduce the family $H_n$ of local hamiltonians we are going
to focus on. To define our family we'll initially fix some two-spin
interaction term $G$ which has bounded norm: $\|G\| \le
\text{const}$. (Note that we can, and will, accommodate next-nearest
neighbour interactions etc.\ by increasing the local dimension of
the spins, i.e.\ by blocking neighbouring spins.) We write the
spectral decomposition of $G$ as $G = \sum_{j=0}^{3}
\lambda_j|\lambda_j\rangle\langle\lambda_j|$. By a trivial rescaling
of the zero point of energy we'll always take
$\lambda_{\text{min}}(G) = \lambda_0 = 0$. Our family $H_n$ of local
quantum systems is then defined by
\begin{equation}
H_n = \sum_{j=0}^{n-1} G_j,
\end{equation}
where $G_j$ is a translate of $G$, i.e., it acts nontrivially on
spins $j$ and $j+1$ as $G$, and as the identity elsewhere.

Now, it is clear that the ground-state energy eigenvalue $E_0\ge
0$ of $H$ will, in general, be strictly positive. In this case, by
using approximate eigenvectors on blocks, it is relatively
straightforward to argue that $E_0$ scales as $E_0\sim c_0 + e_0
n$, where $c_0$ and $e_0$ are constants. Typically $c_0 <0$ and
$e_0> 0$. Because of this scaling it makes sense to talk about the
\emph{ground-state energy density} $E_0/n\rightarrow e_0$. It is
this quantity we wish to bound.

The central part of our argument relies on the Golden-Thompson
inequality \cite{golden:1965, thompson:1965a} (this is Corollary
IX.3.6 in \cite{bhatia:1997a}):
\begin{equation}\label{eq:gt}
\tr(e^{A+B})\le \tr(e^Ae^B),
\end{equation}
where $A$ and $B$ are hermitian operators. To apply the
Golden-Thompson inequality we divide the hamiltonian $H$ into two
pieces:
\begin{equation}\label{eq:hampart}
H = A+B,
\end{equation}
where $A = \sum_{j=0}^{n/2-1}G_{2j}$ and $B = \sum_{j=0}^{n/2-1}
G_{2j+1}$.

\begin{figure}
\includegraphics{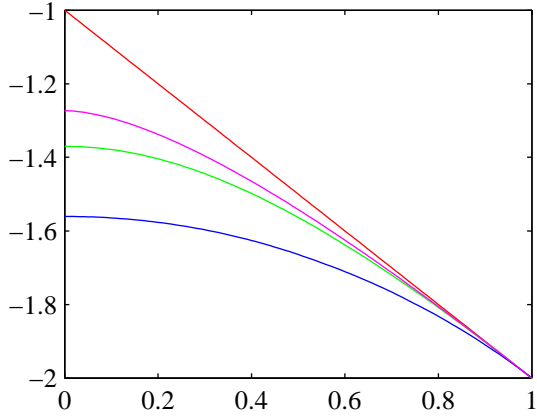}
\caption{Ground-state energy density estimates for $H =
\sum_{j\in\mathbb{Z}} (1+\gamma)\sigma^x_j\otimes \sigma^x_{j+1} +
(1-\gamma)\sigma^y_j\otimes \sigma^y_{j+1}$ versus $\gamma$. The
blue curve is the estimate Eq.~(\ref{eq:gslb}) applied to the basic
even/odd partition Eq.~(\ref{eq:hampart}). The green curve is the
estimate Eq.~(\ref{eq:gslb}) applied to the model where blocks of
$2$ sites have been made. The red curve is mean-field theory. The
magenta curve is the exact ground-state energy density
\cite{lieb:1961a}. The trivial lower bound \cite{anderson:1951a}
that can be derived by diagonalising the interaction term $G$ in
isolation (i.e.\ by pretending each interaction term commutes with
every other one) is $\langle e_0 \rangle \ge -2$ for all $\gamma\in
[0, 1]$.}\label{fig:gsm}
\end{figure}

Our argument works by first applying the Golden-Thompson
inequality Eq.~(\ref{eq:gt}) to the thermal state $e^{-\beta H}$
using the partition Eq.~(\ref{eq:hampart}). This gives us the
inequality
\begin{equation}\label{eq:basicineq}
\mathcal{Z}(\beta) = \tr(e^{-\beta H}) \le \tr(e^{-\beta
A}e^{-\beta B}),
\end{equation}
where $\mathcal{Z}(\beta)$ is the partition function. Next we
study the expression $e^{-\beta A}e^{-\beta B}$. We begin by
writing
\begin{equation}
e^{-\beta G} = \sum_{\alpha, \beta = 0}^3
M(\beta)_{\alpha\beta}\sigma^{\alpha}\otimes \sigma^{\beta},
\end{equation}
where $
\sigma^{\alpha} = \left[ \left(\begin{smallmatrix} 1 & 0 \\
0 & 1
\end{smallmatrix}\right), \left(\begin{smallmatrix} 0 & 1 \\ 1 & 0
\end{smallmatrix}\right), \left(\begin{smallmatrix} 0 & -i \\ i & 0
\end{smallmatrix}\right), \left(\begin{smallmatrix} 1 & 0 \\ 0 & -1
\end{smallmatrix}\right) \right],
$ is the vector of Pauli sigma matrices.

We use this expansion to derive an expression for $e^{-\beta
A}e^{-\beta B}$:
\begin{multline}\label{eq:exprod}
e^{-\beta A}e^{-\beta B} = \sum_{\boldsymbol{\alpha}}
\sum_{\boldsymbol{\alpha}'} M(\beta)_{\alpha_0\alpha_1}\cdots
M(\beta)_{\alpha_{n-2}\alpha_{n-1}}\times
\\ M(\beta)_{\alpha_1'\alpha_2'}\cdots
M(\beta)_{\alpha_{n-1}'\alpha_{0}'}
\sigma^{\boldsymbol{\alpha}}\sigma^{\boldsymbol{\alpha}'},
\end{multline}
$ \sigma^{\boldsymbol{\alpha}} = \sigma^{\alpha_0}\otimes
\sigma^{\alpha_1} \otimes \cdots \sigma^{\alpha_{n-1}}$,
$\alpha_j\in \{0, 1, 2, 3\}$, $0\le j\le n-1$, the \emph{standard
operator basis}.

Substituting the expansion Eq.~(\ref{eq:exprod}) into
Eq.~(\ref{eq:basicineq}) gives us
\begin{equation}
\tr(e^{-\beta H}) \le 2^n\tr(\mathbf{M}(\beta)^n).
\end{equation}
The next step is to bound the partition function from below:
\begin{equation}\label{eq:nextineqy}
e^{-\beta E_0}\le \tr(e^{-\beta H}) \le
2^n\tr(\mathbf{M}(\beta)^n).
\end{equation}
We now observe, thanks to the positivity of
$\tr(\mathbf{M}(\beta)^n)$ for all $n\in \mathbb{N}$, that
\begin{equation}
\tr(\mathbf{M}(\beta)^n) \le D\|\mathbf{M}(\beta)\|_\infty^n,
\end{equation}
where $D$ is the dimension of $\mathbf{M}$. Applying this
inequality to Eq.~(\ref{eq:nextineqy}) and taking logs gives us
\begin{equation}
-\beta E_0 \le \log(D) + n\log(\|2\mathbf{M}(\beta)\|).
\end{equation}
After rearranging we obtain the following inequality for the
ground-state energy density:
\begin{equation}\label{eq:fundam1}
E_0/n \ge -\log(D)/(\beta n) -\log(\|2\mathbf{M}(\beta)\|)/\beta,
\end{equation}
for all $\beta \in [0,\infty)$.

We will be interested in the large-$n$ limit where $n \gg
1/\beta$. In this case we can ignore the first contribution on the
RHS of Eq.~(\ref{eq:fundam1}). Thus we obtain our fundamental
inequality
\begin{equation}\label{eq:gslb}
E_0/n \ge - \inf_{\beta\in (0,\infty)}
\log(\|2\mathbf{M}(\beta)\|)/\beta
\end{equation}

In principle our fundamental inequality Eq.~(\ref{eq:gslb})
provides an analytic lower bound for $E_0/n$. In practice,
however, we need to resort to numerical evaluation of the RHS.
This is an efficient procedure (in $n$) and, thanks to the
continuity of $\log(\|2\mathbf{M}(\beta)\|_\infty)/\beta$ as a
function of $\beta\in(0,\infty)$, provides a certified lower bound
for $E_0/n$.

Can the lower bound Eq.~(\ref{eq:gslb}) be improved? There are at
least two ways to proceed. The first method is to combine $l$
contiguous spins into blocks $\Lambda$ and regard each block
$\Lambda$ as a fundamental spin (of dimension $2^l$). We similarly
block the hamiltonian $H$, which leads to a new nearest-neighbour
hamiltonian for the bigger spins. We can then apply the procedure
we outlined above, albeit using a different operator basis from
standard operator basis. This will lead to a new lower bound. It
is obvious that this procedure cannot lead to a bound which is any
worse: after all, by taking $l = n$ we recover the \emph{exact}
value of $E_0$. The second procedure is to look for decompositions
of $H$ where $H = \sum_{j=0}^{n-1} \widetilde{H}_j$ and where
$\widetilde{H}_j$ has a \emph{larger} minimum eigenvalue. We then
apply the procedure described above using $\widetilde{H}_j$
instead of $H_j$. We'll describe a systematic procedure to obtain
such decompositions in a future paper.

There are several obvious generalisations of our method. The first
generalisation is to calculate $e^{-\beta H_\Lambda}$ for very
large blocks $\Lambda$ of spins via numerical RG methods such as
those described in \cite{zwolak:2004a} and
\cite{verstraete:2004b}. In this case we must put error bars on
the calculated lower bound because of the truncations required by
the methods of \cite{zwolak:2004a} and \cite{verstraete:2004b}.
Applying this technique can give us, in principle, arbitrarily
good lower bounds to the ground-state energy density.

The second generalisation is to quantum spin systems in dimensions
higher than one. In this case it is straightforward to apply our
argument to blocks of spins and thus derive an expression for the
upper bound which resembles the tensor contraction patterns
investigated in \cite{verstraete:2004a} and \cite{murg:2005a}.
Applying the methods described there to approximate such tensor
contraction patterns will allow us to derive lower bounds for the
ground state energy of \emph{finite} higher-dimensional spin
systems.

The third generalisation applies to disordered systems. It is
entirely straightforward to allow the interaction $G$ to vary from
site to site; the method described above applies with essentially
no change to such systems. However, in this case, we need to
average the lower bound over the ensemble of possible
interactions. In principle this can be done analytically for
several models. To illustrate the utility of our lower bound for
disordered systems we apply this technique to the random
antiferromagnetic heisenberg model:
\begin{equation}
H = \sum_{k\in\mathbb{Z}} J_k
\boldsymbol{\sigma}_k\cdot\boldsymbol{\sigma}_{k+1},
\end{equation}
where $J_k$ is a random variable with probability distribution
function $\mu(x)$. Applying our bound Eq.~(\ref{eq:gslb}),
appropriately modified for disordered systems, and averaging over
the measure $d\mu(x)$ gives us the following lower bound for the
expected ground-state energy density:
\begin{equation}
\langle e_0 \rangle \ge -3\langle x\rangle + \log(2)/\beta -\int
d\mu(x)\, \frac{\log(1 + 3e^{-4\beta x})}{\beta},
\end{equation}
for all $\beta\in (0, \infty)$, where $\langle A(x) \rangle = \int
d\mu(x)\, A(x)$. Choosing, for example, $d\mu(x) =
e^{-x^2}/\mathcal{N}$, where $\mathcal{N}$ is a normalisation, and
choosing $x\in[0,\infty)$ provides us with the (numerically
obtained) lower bound $\langle e_0 \rangle \ge -0.833610$. Compare
this lower bound with the lower bound obtained from the Anderson
bound \cite{anderson:1951a}: $\langle e_0 \rangle \ge -1.32934$.

Finally, we say a couple of words about the optimality of our
approach. Consider the interaction term
\begin{equation}
G = -|01\rangle\langle01| + 2|10\rangle \langle 10| + 3|11\rangle
\langle 11|.
\end{equation}
Now the Anderson bound applied to $H= \sum_{j=0}^{n-1} G_j$ yields a
lower bound for the ground-state energy density given by $-1$
whereas the method we've developed here yields the exact answer:
$e_0 = 0$. Even after blocking $m$ spins the Anderson lower bound is
still only $-1/m$.

The decomposition Eq.~(\ref{eq:exprod}) and subsequent derivation
are strongly reminiscent of DMRG-type methods based on \emph{matrix
product states} (MPS) \cite{porras:2005a}. It is certainly true that
methods for the ground-state energy based on MPS reduce to
variational problems over a restricted class of vectors in hilbert
space and thus must have a well defined lagrangian dual which would,
at least in principle, provide lower bounds for the ground-state
energy density. (Further investigation of this dual problem will be
reported elsewhere). It is currently unclear what, if any,
connection our method has to this dual problem.

\begin{acknowledgements}
I would like to thank Tony Short for several helpful discussions.
I am grateful to the EU for support for this research under the
IST project RESQ and also to the UK EPSRC through the grant
QIPIRC.
\end{acknowledgements}


\begin{thebibliography}{18}
\expandafter\ifx\csname
natexlab\endcsname\relax\def\natexlab#1{#1}\fi
\expandafter\ifx\csname bibnamefont\endcsname\relax
  \def\bibnamefont#1{#1}\fi
\expandafter\ifx\csname bibfnamefont\endcsname\relax
  \def\bibfnamefont#1{#1}\fi
\expandafter\ifx\csname citenamefont\endcsname\relax
  \def\citenamefont#1{#1}\fi
\expandafter\ifx\csname url\endcsname\relax
  \def\url#1{\texttt{#1}}\fi
\expandafter\ifx\csname urlprefix\endcsname\relax\def\urlprefix{URL
}\fi \providecommand{\bibinfo}[2]{#2}
\providecommand{\eprint}[2][]{\url{#2}}

\bibitem[{\citenamefont{Sachdev}(1999)}]{sachdev:1999a}
\bibinfo{author}{\bibfnamefont{S.}~\bibnamefont{Sachdev}},
  \emph{\bibinfo{title}{Quantum phase transitions}}
  (\bibinfo{publisher}{Cambridge University Press},
  \bibinfo{address}{Cambridge}, \bibinfo{year}{1999}).

\bibitem[{\citenamefont{Kitaev et~al.}(2002)\citenamefont{Kitaev, Shen, and
  Vyalyi}}]{kitaev:2002a}
\bibinfo{author}{\bibfnamefont{A.~Y.} \bibnamefont{Kitaev}},
  \bibinfo{author}{\bibfnamefont{A.~H.} \bibnamefont{Shen}}, \bibnamefont{and}
  \bibinfo{author}{\bibfnamefont{M.~N.} \bibnamefont{Vyalyi}},
  \emph{\bibinfo{title}{Classical and quantum computation}},
  vol.~\bibinfo{volume}{47} of \emph{\bibinfo{series}{Graduate Studies in
  Mathematics}} (\bibinfo{publisher}{American Mathematical Society},
  \bibinfo{address}{Providence, RI}, \bibinfo{year}{2002}).

\bibitem[{\citenamefont{Kempe et~al.}(2004)\citenamefont{Kempe, Kitaev, and
  Regev}}]{kempe:2004a}
\bibinfo{author}{\bibfnamefont{J.}~\bibnamefont{Kempe}},
  \bibinfo{author}{\bibfnamefont{A.}~\bibnamefont{Kitaev}}, \bibnamefont{and}
  \bibinfo{author}{\bibfnamefont{O.}~\bibnamefont{Regev}}, in
  \emph{\bibinfo{booktitle}{FSTTCS 2004: Foundations of software technology and
  theoretical computer science}} (\bibinfo{publisher}{Springer},
  \bibinfo{address}{Berlin}, \bibinfo{year}{2004}), vol. \bibinfo{volume}{3328}
  of \emph{\bibinfo{series}{Lecture Notes in Comput. Sci.}}, pp.
  \bibinfo{pages}{372--383}, \eprint{quant-ph/0406180}.

\bibitem[{\citenamefont{Oliveira and Terhal}(2005)}]{oliveira:2005a}
\bibinfo{author}{\bibfnamefont{R.}~\bibnamefont{Oliveira}} \bibnamefont{and}
  \bibinfo{author}{\bibfnamefont{B.~M.} \bibnamefont{Terhal}}
  (\bibinfo{year}{2005}), \eprint{quant-ph/0504050}.

\bibitem[{\citenamefont{Schollw{\"o}ck}(2005)}]{schollwoeck:2005a}
\bibinfo{author}{\bibfnamefont{U.}~\bibnamefont{Schollw{\"o}ck}},
  \bibinfo{journal}{Rev. Modern Phys.} \textbf{\bibinfo{volume}{77}},
  \bibinfo{pages}{259} (\bibinfo{year}{2005}), \eprint{cond-mat/0409292}.

\bibitem[{\citenamefont{Verstraete and Cirac}(2004)}]{verstraete:2004a}
\bibinfo{author}{\bibfnamefont{F.}~\bibnamefont{Verstraete}} \bibnamefont{and}
  \bibinfo{author}{\bibfnamefont{J.~I.} \bibnamefont{Cirac}}
  (\bibinfo{year}{2004}), \eprint{cond-mat/0407066}.

\bibitem[{\citenamefont{Anderson}(1951)}]{anderson:1951a}
\bibinfo{author}{\bibfnamefont{P.~W.} \bibnamefont{Anderson}},
  \bibinfo{journal}{Phys. Rev.} \textbf{\bibinfo{volume}{83}},
  \bibinfo{pages}{1260} (\bibinfo{year}{1951}).

\bibitem[{\citenamefont{Wittmann and Stolze}(1993)}]{wittmann:1993a}
\bibinfo{author}{\bibfnamefont{T.}~\bibnamefont{Wittmann}} \bibnamefont{and}
  \bibinfo{author}{\bibfnamefont{J.}~\bibnamefont{Stolze}},
  \bibinfo{journal}{Phys. Rev. B} \textbf{\bibinfo{volume}{48}},
  \bibinfo{pages}{3479} (\bibinfo{year}{1993}).

\bibitem[{\citenamefont{Bratteli and Robinson}(1987)}]{bratteli:1987a}
\bibinfo{author}{\bibfnamefont{O.}~\bibnamefont{Bratteli}} \bibnamefont{and}
  \bibinfo{author}{\bibfnamefont{D.~W.} \bibnamefont{Robinson}},
  \emph{\bibinfo{title}{Operator algebras and quantum statistical mechanics.
  1}}, Texts and Monographs in Physics (\bibinfo{publisher}{Springer-Verlag},
  \bibinfo{address}{New York}, \bibinfo{year}{1987}), \bibinfo{edition}{2nd}
  ed.

\bibitem[{\citenamefont{Bratteli and Robinson}(1997)}]{bratteli:1997a}
\bibinfo{author}{\bibfnamefont{O.}~\bibnamefont{Bratteli}} \bibnamefont{and}
  \bibinfo{author}{\bibfnamefont{D.~W.} \bibnamefont{Robinson}},
  \emph{\bibinfo{title}{Operator algebras and quantum statistical mechanics.
  2}}, Texts and Monographs in Physics (\bibinfo{publisher}{Springer-Verlag},
  \bibinfo{address}{Berlin}, \bibinfo{year}{1997}), \bibinfo{edition}{2nd} ed.

\bibitem[{\citenamefont{Golden}(1965)}]{golden:1965}
\bibinfo{author}{\bibfnamefont{S.}~\bibnamefont{Golden}},
  \bibinfo{journal}{Phys. Rev.} \textbf{\bibinfo{volume}{137}},
  \bibinfo{pages}{B1127} (\bibinfo{year}{1965}).

\bibitem[{\citenamefont{Thompson}(1965)}]{thompson:1965a}
\bibinfo{author}{\bibfnamefont{C.~J.} \bibnamefont{Thompson}},
  \bibinfo{journal}{J. Mathematical Phys.} \textbf{\bibinfo{volume}{6}},
  \bibinfo{pages}{1812} (\bibinfo{year}{1965}).

\bibitem[{\citenamefont{Bhatia}(1997)}]{bhatia:1997a}
\bibinfo{author}{\bibfnamefont{R.}~\bibnamefont{Bhatia}},
  \emph{\bibinfo{title}{Matrix analysis}}
  (\bibinfo{publisher}{Springer-Verlag}, \bibinfo{address}{New York},
  \bibinfo{year}{1997}).

\bibitem[{\citenamefont{Lieb et~al.}(1961)\citenamefont{Lieb, Schultz, and
  Mattis}}]{lieb:1961a}
\bibinfo{author}{\bibfnamefont{E.}~\bibnamefont{Lieb}},
  \bibinfo{author}{\bibfnamefont{T.}~\bibnamefont{Schultz}}, \bibnamefont{and}
  \bibinfo{author}{\bibfnamefont{D.}~\bibnamefont{Mattis}},
  \bibinfo{journal}{Ann. Physics} \textbf{\bibinfo{volume}{16}},
  \bibinfo{pages}{407} (\bibinfo{year}{1961}).

\bibitem[{\citenamefont{Zwolak and Vidal}(2003)}]{zwolak:2004a}
\bibinfo{author}{\bibfnamefont{M.}~\bibnamefont{Zwolak}} \bibnamefont{and}
  \bibinfo{author}{\bibfnamefont{G.}~\bibnamefont{Vidal}},
  \bibinfo{journal}{Phys. Rev. Lett.} \textbf{\bibinfo{volume}{93}},
  \bibinfo{pages}{207205} (\bibinfo{year}{2003}), \eprint{cond-mat/0406440}.

\bibitem[{\citenamefont{Verstraete et~al.}(2004)\citenamefont{Verstraete,
  Garc{\'\i}a-Ripoll, and Cirac}}]{verstraete:2004b}
\bibinfo{author}{\bibfnamefont{F.}~\bibnamefont{Verstraete}},
  \bibinfo{author}{\bibfnamefont{J.~J.} \bibnamefont{Garc{\'\i}a-Ripoll}},
  \bibnamefont{and} \bibinfo{author}{\bibfnamefont{J.~I.} \bibnamefont{Cirac}},
  \bibinfo{journal}{Phys. Rev. Lett.} \textbf{\bibinfo{volume}{93}},
  \bibinfo{pages}{207204} (\bibinfo{year}{2004}), \eprint{quant-ph/0406426}.

\bibitem[{\citenamefont{Murg et~al.}(2005)\citenamefont{Murg, Verstraete, and
  Cirac}}]{murg:2005a}
\bibinfo{author}{\bibfnamefont{V.}~\bibnamefont{Murg}},
  \bibinfo{author}{\bibfnamefont{F.}~\bibnamefont{Verstraete}},
  \bibnamefont{and} \bibinfo{author}{\bibfnamefont{J.~I.} \bibnamefont{Cirac}},
  \bibinfo{journal}{Phys. Rev. Lett.} \textbf{\bibinfo{volume}{95}},
  \bibinfo{pages}{057206} (\bibinfo{year}{2005}), \eprint{quant-ph/0501493}.

\bibitem[{\citenamefont{Porras et~al.}(2005)\citenamefont{Porras, Verstraete,
  and Cirac}}]{porras:2005a}
\bibinfo{author}{\bibfnamefont{D.}~\bibnamefont{Porras}},
  \bibinfo{author}{\bibfnamefont{F.}~\bibnamefont{Verstraete}},
  \bibnamefont{and} \bibinfo{author}{\bibfnamefont{J.~I.} \bibnamefont{Cirac}}
  (\bibinfo{year}{2005}), \eprint{quant-ph/0504717}.

\end{thebibliography}
\end{document}